# Lithium-Metal Batteries Using Sustainable Electrolyte Media and Various Cathode Chemistries

Vittorio Marangon[a], Luca Minnetti[b], Matteo Adami[a], Alberto Barlini[a], Jusef Hassoun[a,b,*]

[a] *University of Ferrara, Department of Chemical and Pharmaceutical Sciences, Via Fossato di Mortara 17, 44121, Ferrara, Italy.*

[b] *Graphene Labs, Istituto Italiano di Tecnologia, Via Morego 30 – 16163 Genova.*

Corresponding Author: jusef.hassoun@unife.it, jusef.hassoun@iit.it

**Abstract**

Lithium-metal batteries employing concentrated glyme-based electrolytes and different cathode chemistries are herein evaluated in view of a safe use of the highly energetic alkali-metal anode. Indeed, diethylene-glycol dimethyl-ether (DEGDME) and triethylene-glycol dimethyl-ether (TREGDME) dissolving lithium bis(trifluoromethanesulfonyl)imide (LiTFSI) and lithium nitrate (LiNO$_3$) in concentration approaching the solvents saturation limit are used in lithium batteries employing either a conversion sulfur-tin composite (S:Sn 80:20 w/w) or a Li$^+$ (de-)insertion LiFePO$_4$ cathode. Cyclic voltammetry (CV) and electrochemical impedance spectroscopy (EIS) clearly show the suitability of the concentrated electrolytes in terms of process reversibility and low interphase resistance, particularly upon a favorable activation. Galvanostatic measurements performed in the lithium-sulfur (Li/S) batteries reveal promising capacities at room temperature (25 °C) and a value as high as 1300 mAh g$_S^{-1}$ for DEGDME-based electrolyte at 35 °C. On the other hand, the lithium-LiFePO$_4$ (Li/LFP) cells exhibit satisfactory cycling behavior, in particular when employing an additional reduction step at low voltage cutoff (i.e., 1.2 V) during the first discharge to consolidate the solid electrolyte interphase (SEI). This procedure allows a coulombic efficiency near 100 %, a capacity approaching 160 mAh g$^{-1}$ and relevant retention particularly for the cell using TREGDME-based electrolyte. Therefore, this work suggests the use of concentrated glyme-based electrolytes, the fine tuning of the operative conditions, and the careful selection of active materials chemistry as significant steps to achieve practical and safe lithium-metal batteries.

**Introduction**

Li-ion battery powers a wide array of electronic devices, from portable systems such as laptops and smartphones, to hybrid (HEVs) and fully electric vehicles (EVs).[1,2] The research on Li-ion batteries has led to the achievement of a remarkable energy density, i.e., 260 Wh kg$^{-1}$, and a long cycle life.[3,4] However, a raising demand for energy with the purpose of extending the driving range of EVs renewed the interest in the metallic lithium anode, which offers a high theoretical capacity (3860 mAh g$^{-1}$) and the lowest redox potential (-3.04 V *vs.* SHE) among the various electrodes proposed as the battery anode.[5] Despite the various advantages, the application of the lithium in a rechargeable battery was so far hindered by the formation of dendritic structures due to heterogeneous deposition of lithium at the metal surface during charge that can lead to short circuits and hazards.[6] The most relevant solutions proposed to overcome this challenging issue and ensure efficient and safe discharge-charge cycling of the battery were represented by the addition to the electrolyte of sacrificial agents such as the lithium nitrate (LiNO$_3$) that can be reduced at the lithium surface to protect the metallic anode by the formation of suitable solid electrolyte interphase (SEI) film.[7–9] A further relevant breakthrough was achieved by the replacement of carbonate-based solvents with more stable and less volatile polyethylene oxides or end-capped glymes (CH$_3$O(CH$_2$CH$_2$O)$_n$CH$_3$).[10–15] Remarkable improvement of the safety content of the cell was furthermore obtained by increasing the salt concentration, in particular using the glyme-based electrolytes, in order to decrease the flammability and the volatility, holding at the same time long cycle life and high coulombic efficiency.[16–19] In this regard, in our previous study we have characterized the chemical and electrochemical properties of highly concentrated di- and triglyme-based electrolytes employing the conductive salt lithium bis(trifluoromethanesulfonyl)imide (LiTFSI) and LiNO$_3$ in concentrations approaching the solvent saturation limit.[20] The study focused on the performance of the new electrolyte media in Li/O$_2$ battery, particularly in terms of electrode/electrolyte interphase effects on the cycling behavior of the cell. The data of the above research principally suggested the triglyme-based electrolyte as promising candidate for application in Li/O$_2$ cell due to its unique properties including a remarkably low volatility and enhanced interphase stability,[20] thus in agreement with other literature papers.[21,22] Indeed, according to X-ray photoelectron spectroscopy (XPS) and electrochemical impedance spectroscopy (EIS), the enhanced characteristics the triglyme-based electrolyte compared to diglyme-based one in Li/O$_2$ cell have been principally attributed to the formation of a stable SEI, in particular on the Li metal.[20]

Herein, we originally extend the investigation of these highly concentrated electrolyte media to different cathode chemistries which can be employed in new configurations of lithium-metal battery, that is, the ones using the conversion electrochemical process related to sulfur[23,24] and the Li$^+$ (de-)insertion mechanism associated to a LiFePO$_4$ olivine cathode.[3,25] Therefore, the present study focuses on the electrochemical performances of the new electrolytes in advanced lithium cells using the high-performances, sulfur-tin composite with low amount of electrochemically-inactive conductive metal (S:Sn 80:20 w/w)[26] and the advanced carbon-coated LiFePO$_4$ cathode.[27] The results of the study may actually shed light on possible applications of the highly concentrated glyme-based electrolyte for achieving new rechargeable batteries with remarkable safety content using the highly energetic, yet challenging lithium-metal anode.

**Experimental section**

*Materials*

Lithium bis(trifluoromethanesulfonyl)imide (LiTFSI, Sigma-Aldrich) and lithium nitrate (LiNO$_3$, Sigma-Aldrich) salts were dissolved in diethylene-glycol dimethyl-ether (DEGDME, CH$_3$O(CH$_2$CH$_2$O)$_2$CH$_3$, Sigma-Aldrich) and triethylene-glycol dimethyl-ether (TREGDME, CH$_3$O(CH$_2$CH$_2$O)$_3$CH$_3$, Sigma-Aldrich) solvents at room temperature overnight under magnetic stirring inside an Ar-filled glovebox (MBraun, O$_2$ and H$_2$O content below 1 ppm). The final concentration of each salt was of 1.5 mol kg$_{solvent}^{-1}$ in DEGDME and 2 mol kg$_{solvent}^{-1}$ in TREGDME, that is, amounts approaching the saturation limit of the solvents. Prior to using, LiTFSI and LiNO$_3$ were dried at 110 °C for 24 hours under vacuum, while DEGDME and TREGDME solvents were dried under molecular sieves (3 Å, Sigma-Aldrich) until a water content lower than 10 ppm was achieved as measured by 899 Karl Fischer Coulometer, Metrohm. The highly concentrated electrolyte solutions are subsequently indicated as DEGDME_HCE and TREGDME_HCE. The analyses of the chemical and electrochemical properties of the electrolytes are reported in a previous work.[20]

The synthesis of the S:Sn 80:20 powder was achieved in a previous work through physical mixing and melting process of elemental sulfur (80 % wt., ≥99.5%, Riedel-de Haën) and nanometric tin powder (20 % wt., <150 nm, Sigma-Aldrich, ≥99% trace metal basis) at 120 °C,[26] while the LiFePO$_4$ (LFP) material was developed by Advanced Lithium Electrochemistry (Aleees Taiwan, Model A1100) and characterized by a carbon content of about 5%.[27]

*Electrochemical measurements*

The electrochemical tests were carried out in CR2032 coin-type cells assembled in Ar-filled glovebox (MBraun, O$_2$ and H$_2$O content below 1 ppm) by employing a 14 mm diameter lithium disk as anode. The S:Sn 80:20 and LFP electrodes were obtained by solvent casting of the active materials on either a porous carbon-cloth foil (GDL, ELAT 1400, MTI Corp.) or an aluminum current collector, respectively. The cathodes were separated from the lithium anode by a 16 mm Celgard foil soaked with the electrolyte solution (either DEGDME_HCE or TREGDME_HCE, see below related amounts) in the Li/S:Sn 80:20 cells, while by two GF/A glass fiber Whatman 16 mm disks soaked with the electrolyte solution (either DEGDME_HCE or TREGDME_HCE) in the Li/LFP cells.

Cyclic voltammetry (CV) was performed at a scan rate of 0.1 mV s$^{-1}$ in the 1.8 – 2.8 V *vs* Li$^+$/Li potential range for S:Sn 80:20 electrode and in the 2.7 – 3.9 V *vs* Li$^+$/Li potential range for LFP electrode. Electrochemical impedance spectra were collected at the open circuit voltage (OCV) condition of the cell, as well as after 1, 5 and 10 CV cycles, and analyzed through the non-linear least squares (NLLS) fitting method via the Boukamp software ($\chi^2$ values of the order of 10$^{-4}$ or lower).[28,29] EIS was performed by applying to the cells an alternate voltage signal with amplitude of 10 mV within the frequency range from 500 kHz to 100 mHz. All the CV and EIS measurements were performed by using a VersaSTAT MC Princeton Applied Research (PAR, AMETEK) instrument. The Li/S:Sn 80:20 cells were tested through galvanostatic cycling measurements carried out at the constant current rate of C/5 at 25 °C and 35 °C and of 1C at 35 °C (1C = 1675 mA g$_S^{-1}$). The cells cycled at the current of C/5 employed 60 µl of electrolyte solution and a voltage range of 1.9 – 2.8 V, while voltage limits of 1.6 and 2.8 V and an optimized electrolyte/sulfur ratio of 20 µl mg$_S^{-1}$ were adopted for the cells tested at 1C. The galvanostatic cycling measurements of the Li/LFP cells were performed at the constant current rate of C/5 (1C = 170 mA g$^{-1}$) at room temperature (25 °C), in a 2.7 – 3.9 V voltage range. Additional tests were carried out at C/5 and C/3 by exploiting a voltage range between 1.2 and 3.9 V during the first charge-discharge cycle and between 2.7 and 3.9 V for the subsequent ones. All the galvanostatic cycling measurements were performed by using a MACCOR series 4000 battery test system.

**Results and Discussion**

The DEGDME_HCE and TREGDME_HCE are initially exploited in lithium cell using the S:Sn 80:20 cathode by means of CV coupled with EIS as reported in Figure 1. The corresponding voltammograms (Fig. 1a, c) show the typical profiles expected for the reversible multiple-step Li/S electrochemical process consisting of a first cycle with a different shape compared to the subsequent ones which are well overlapped into various peaks centered at about 2.4 and below 2.0 V $vs$ Li$^+$/Li during cathodic scan and merged between 2.3 and 2.5 V $vs$ Li$^+$/Li during the anodic one.[23] The above peaks correspond to the reduction with formation of soluble polysulfides (Li$_2$S$_x$ with x ≥ 6 at 2.4 V, and Li$_2$S$_x$ with 6 > x > 2 below 2.0 V) during discharge, and oxidation back to sulfur during charge process.[24] Furthermore, the difference between first and subsequent cycles is well justified by the EIS performed at the OCV and after 1, 5 and 10 voltammetry cycles for the cells using DEGDME_HCE (Fig. 1b) and TREGDME_HCE (Fig. 1d). The corresponding Nyquist plots suggest the activation process typical of Li/S cells using a suitable electrolyte associated with the consolidation of a favorable electrode/electrolyte interphase by the ongoing of the electrochemical process.[26] This Li/S activation process has been attributed in a previous paper to micro-structural modifications of the electrode that allow an enhanced electric contact between sulfur and the conductive carbon support, and lead to an improved conductivity of the electrode/interphase.[30] Indeed, the EIS evidences a remarkable decrease of the cell impedance from values higher than 100 Ω at the OCV (see insets of Fig. 1b and d) to values of the order of 10 Ω for DEGDME_HCE (Fig. 1b) and 20 Ω for TREGDME_HCE (Fig. 1d). An exhaustive summary of the results of the NLLS analyses performed on the Nyquist plots of Fig. 1b and Fig. 1d is reported in Table 1.[28,29] Notably, both CV and EIS indicate differences between the Li/S cells using DEGDME_HCE and TREGDME_HCE: the former shows smoother less polarized peaks and a slightly lower steady state impedance with respect to the latter (compare Fig. 1a and inset of Fig.1b with Fig. 1c and inset of Fig. 1d, respectively). These differences may be likely associated with favorable effects on the Li/S electrochemical process promoted by the lower solvent viscosity (0.94 g ml$^{-1}$)[31] and higher conductivity (3.3×10$^{-3}$ S cm$^{-1}$) of DEGDME_HCE compared to the TREGDME_HCE (0.98 g ml$^{-1}$ and 8.9×10$^{-4}$ S cm$^{-1}$, respectively)[31] at room temperature, as reported in our previous study.[20]

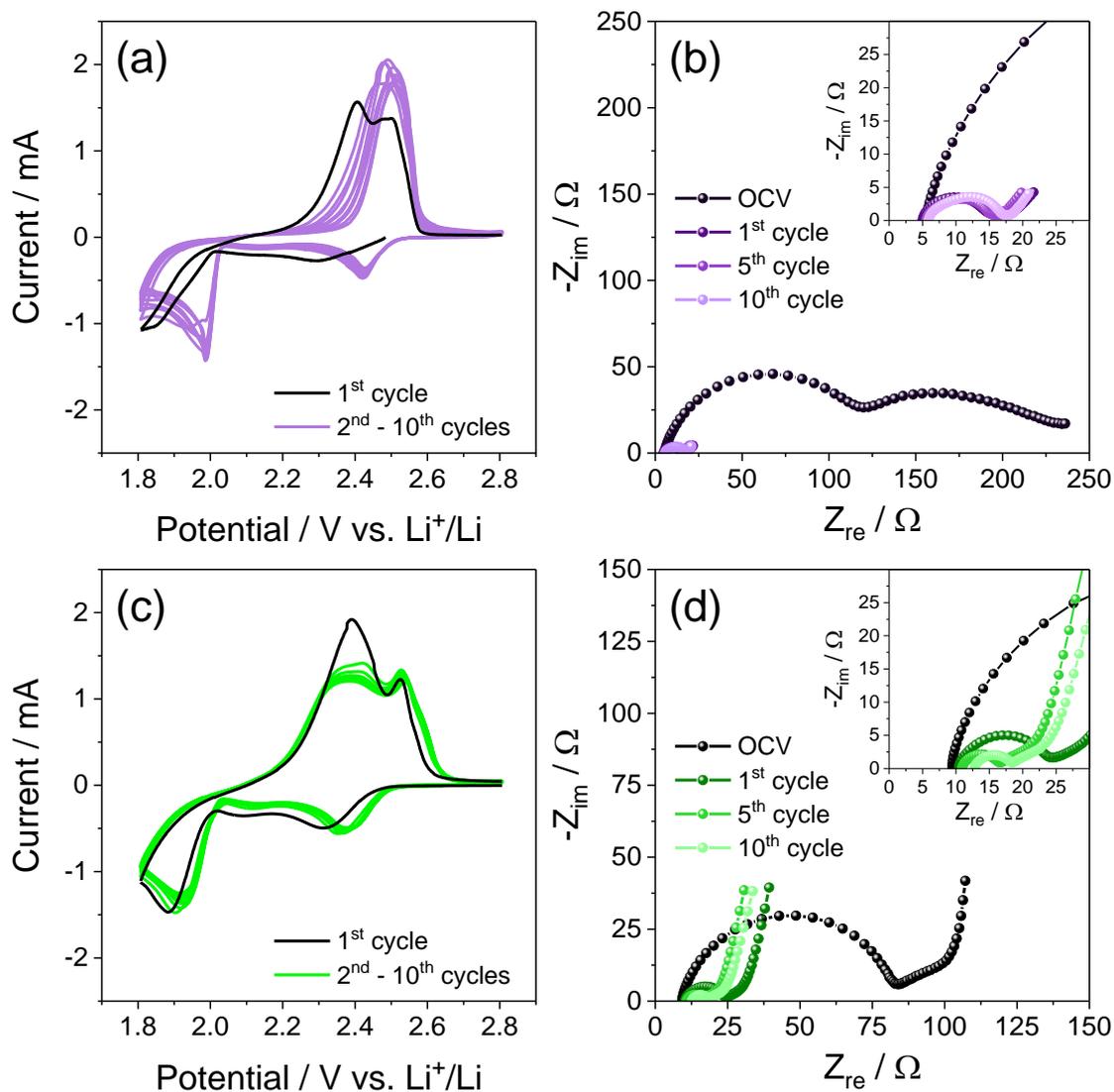

**Figure 1.** **(a,c)** Cyclic voltammetry (CV) and **(b,d)** electrochemical impedance spectroscopy (EIS) measurements performed on Li/electrolyte/S:Sn 80:20 cells employing either **(a,b)** DEGDME_HCE or **(c,d)** TREGDME_HCE. CV potential range: 1.8 – 2.8 V *vs.* Li$^+$/Li; scan rate: 0.1 mV s$^{-1}$. EIS carried out at the OCV of the cells and after 1, 5 and 10 voltammetry cycles (inset reports magnification); frequency range: 500 kHz – 100 mHz; signal amplitude: 10 mV.

**Table 1.** NLLS analyses performed on the Nyquist plots reported in Figure 1b and d recorded upon CV measurements of Li/electrolyte/S:Sn 80:20 cells employing either DEGDME_HCE (Fig. 1b) or TREGDME_HCE (Fig. 1d).[28,29]

| Electrolyte | Cell condition | Circuit | $R_1$ [Ω] | $R_2$ [Ω] | $R_1 + R_2$ [Ω] | $\chi^2$ |
|---|---|---|---|---|---|---|
| DEGDME_HCE | OCV | $R_e(R_1Q_1)(R_2Q_2)$ | 103 ± 2 | 129 ± 3 | 232 ± 4 | $1\times10^{-4}$ |
| | 1 cycle | $R_e(R_1Q_1)(R_2Q_2)Q_3$ | 9.9 ± 0.1 | 3.4 ± 0.2 | 13.3 ± 0.2 | $2\times10^{-5}$ |
| | 5 cycles | $R_e(R_1Q_1)(R_2Q_2)Q_3$ | 10.1 ± 0.1 | 0.8 ± 0.1 | 10.9 ± 0.1 | $3\times10^{-5}$ |
| | 10 cycles | $R_e(R_1Q_1)(R_2Q_2)Q_3$ | 9.8 ± 0.5 | 1.5 ± 0.5 | 11.3 ± 0.7 | $6\times10^{-5}$ |
| TREGDME_HCE | OCV | $R_e(R_1Q_1)(R_2Q_2)Q_3$ | 73.3 ± 0.4 | 27 ± 3 | 101 ± 3 | $5\times10^{-5}$ |
| | 1 cycle | $R_e(R_1Q_1)(R_2Q_2)Q_3$ | 13.1 ± 0.3 | 9.0 ± 1.5 | 22.1 ± 1.5 | $4\times10^{-5}$ |
| | 5 cycles | $R_e(R_1Q_1)(R_2Q_2)Q_3$ | 5.3 ± 0.1 | 7.2 ± 0.3 | 12.5 ± 0.3 | $2\times10^{-5}$ |
| | 10 cycles | $R_e(R_1Q_1)(R_2Q_2)Q_3$ | 5.4 ± 0.1 | 6.8 ± 0.3 | 12.2 ± 0.3 | $2\times10^{-5}$ |

Figure 2 displays the performance of the DEGDME_HCE (Fig. 2a, b) and TREGDME_HCE (Fig. 2c, d) in a Li/S:Sn 80:20 cell, cycled at the constant rate of C/5 (1C = 1675 mA g$_s^{-1}$) at 25 and 35 °C. The selected voltage profiles related to the steady state of the cell (Fig. 2a, c) reveal the characteristic response of a Li/S battery, in agreements with CVs of Fig. 1, where the two distinct discharge plateaus around 2.4 and 2.0 V ascribed to the formation of long chain lithium polysulfides (Li$_2$S$_x$ with x ≥ 6) and short chain ones (Li$_2$S$_x$ with 6 > x > 2), respectively, are reversed into two charge plateaus above 2.3 V.[30,32] Furthermore, the figure shows for both DEGDME_HCE (Fig. 2a) and TREGDME_HCE (Fig. 2c) a relatively high polarization at the room temperature (25 °C), in particular for the latter electrolyte, which leads to steady state specific capacities of about 800 mAh g$_s^{-1}$ and 340 mAh g$_s^{-1}$, respectively. The poor response at the room temperature of the Li/S cells is most likely due to the hindered kinetics of the insulating sulfur by the high viscosity of the concentrated electrolytes, which is particularly relevant in the case of the TREGDME_HCE due to its longer ether-chain, higher lithium salts concentration (see experimental section), and consequent higher viscosity compared to DEGDME_HCE.[20,33] In order to favor the electrochemical kinetics and achieve better performances, subsequent galvanostatic cycling tests were performed on the Li/S:Sn 80:20 cells at a higher temperature, that is, 35 °C, by employing the same C-rate of C/5 (1C = 1675 mA g$_S^{-1}$). Advantageously, the increase of temperature leads to higher capacity values and to lower polarization for both DEGDME_HCE (Fig. 2a) and TREGDME_HCE (Fig. 2c), as expected by the decrease of the electrolytes viscosity and the concomitant rise of their Li$^+$ ions conductivity.[20]

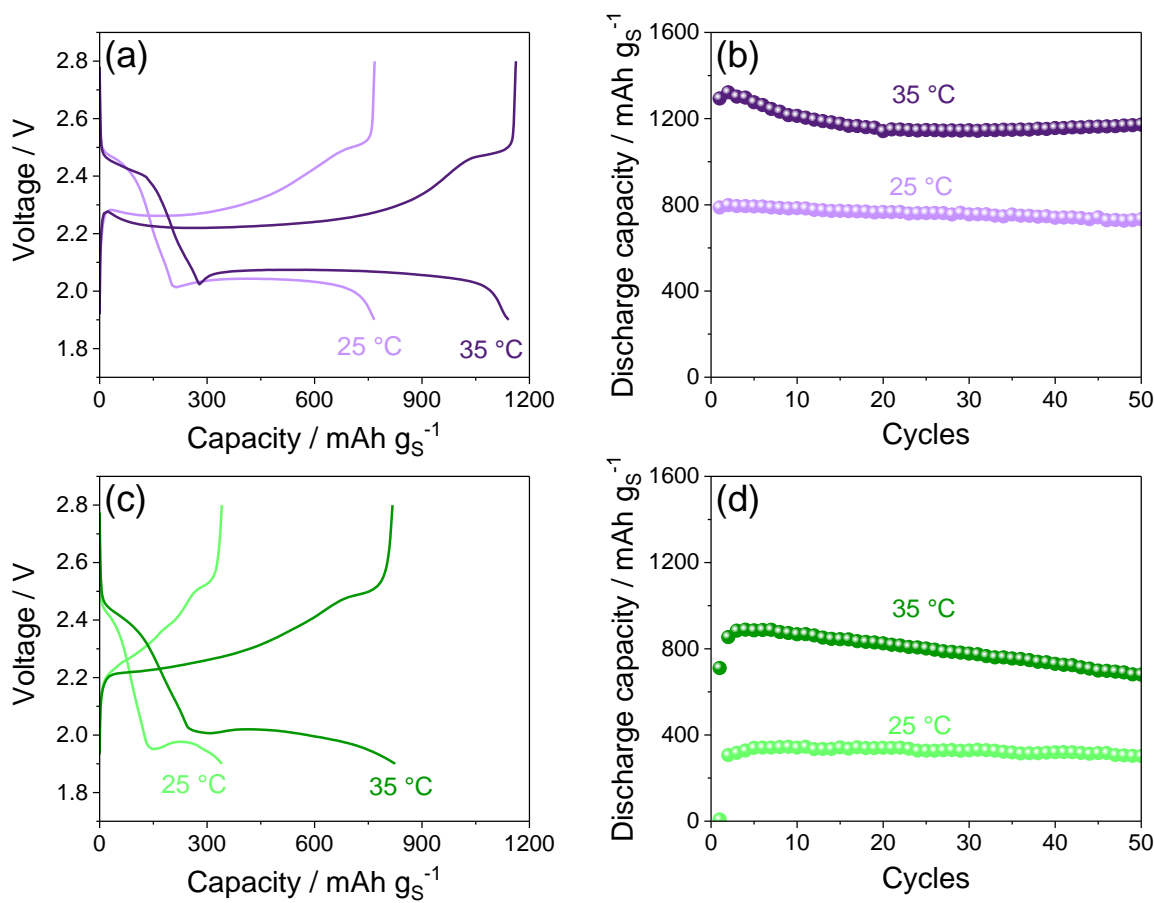

**Figure 2. (a,c)** Selected voltage profiles and **(b,d)** corresponding cycling trends at 25 and 35 °C of Li/electrolyte/S:Sn 80:20 cells employing either **(a,b)** DEGDME_HCE or **(c,d)** TREGDME_HCE. The cells are galvanostatically cycled using a voltage range between 1.9 and 2.8 V at the constant current rate of C/5 (1C = 1675 mA g$_S^{-1}$).

In particular, the cycling trends reported in Figure 2b and d reveal that the cell using DEGDME_HCE delivers maximum specific capacity of about 1320 mAh g$_S^{-1}$ (Fig. 2b), while the one exploiting TREGDME_HCE exhibits values as high as 890 mAh g$_S^{-1}$ (Fig. 2d). Furthermore, the cell using DEGDME_HCE shows an excellent capacity retention with values ranging from 92 % at the room temperature to 90 % at 35 °C, over the 50 cycles of the tests (Fig. 2b), while lower but still satisfactory values of 88 % at the room temperature and 77 % at 35 °C, are observed for the cell using the more viscose TREGDME_HCE (Fig. 2d).

A further application of the studied electrolytes is exploited by lithium-metal cells using an olivine-structured, (de-)insertion LiFePO$_4$ (LFP) olivine cathode[3] which is identified by literature works as a promising candidate for lithium-metal batteries using concentrated electrolytes.[34,35] A combined study using CV and EIS is therefore performed and reported in Figure 3, analogously to the investigation provided for the sulfur-based electrode (compare with Fig. 1). The voltammograms of the cells using DEGDME_HCE and TREGDME_HCE (Fig. 3a and c, respectively) show the characteristic profile centered at about 3.45 V *vs* Li$^+$/Li associated with the de-insertion of Li from the LiFePO$_4$ during the anodic scan and its insertion back into the olivine structure during the cathodic scan[3,25] The first CV cycle shows a charge/discharge polarization of about 0.3 V *vs* Li$^+$/Li for DEGDME_HCE (Fig. 3a, black curve) and of about 0.4 V *vs* Li$^+$/Li for TREGDME_HCE (Fig. 3c, black curve). This relatively high polarization may be ascribed to a not yet completely optimized electrode/electrolyte interphase of the LFP electrode using the highly concentrated electrolytes.[10] Furthermore, the subsequent cycles reveal for the two electrolytes a certain improvement with favorable decrease of the above mentioned polarization, leading to a shift of about 0.1 V *vs* Li$^+$/Li of the cathodic and the anodic peaks. This enhancement is likely justified by the EIS Nyquist plots reported in Fig. 3b for DEGDME_HCE and in Fig. 3d for TREGDME_HCE and by the results of the corresponding NLLS analyses listed in Table 2.[28,29] Indeed, the data indicate for the two electrolytes a series of semicircles and lines ascribed to SEI layers, charge transfer processes and diffusion phenomena occurring in the lithium cell at the various frequencies, with an overall initial resistance of about 300 Ω for DEGDME_HCE and 150 Ω for TREGDME_HCE decreasing to about 90 Ω and 80 Ω, respectively, upon the 10 CV cycles taken under consideration. This likely indicates a partial dissolution or a modification of the pristine SEI layer formed at the electrodes surface by the ongoing of the cycling, that can initially favor the kinetics of the electrochemical processes.[7,36]

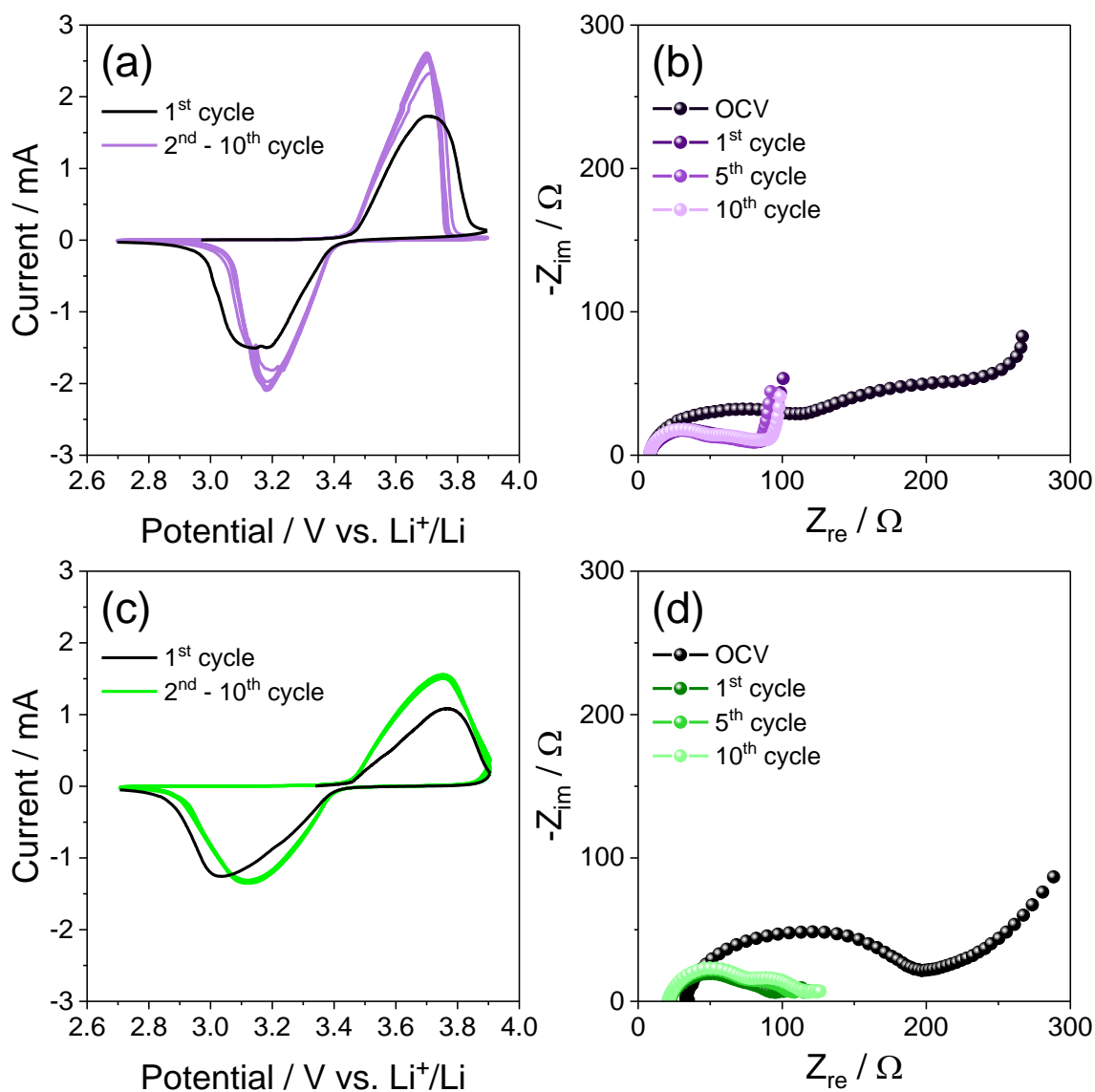

**Figure 3. (a,c)** Cyclic voltammetry (CV) and **(b,d)** electrochemical impedance spectroscopy (EIS) measurements performed on Li/electrolyte/LFP cells employing either **(a,b)** DEGDME_HCE or **(c,d)** TREGDME_HCE. CV potential range: 2.7 – 3.9 V *vs.* Li$^+$/Li; scan rate: 0.1 mV s$^{-1}$. EIS carried out at the OCV of the cells and after 1, 5 and 10 voltammetry cycles; frequency range: 500 kHz – 100 mHz; signal amplitude: 10 mV.

**Table 2.** NLLS analyses performed on the Nyquist plots reported in Figure 3b and d recorded upon CV measurements of Li/electrolyte/LFP cells employing either DEGDME_HCE (Fig. 3b) or TREGDME_HCE (Fig. 3d).[28,29]

| Electrolyte | Cell condition | Circuit | $R_1$ [Ω] | $R_2$ [Ω] | $R_1 + R_2$ [Ω] | $\chi^2$ |
|---|---|---|---|---|---|---|
| DEGDME_HCE | OCV | $R_e(R_1Q_1)(R_2Q_2)Q_3$ | 99 ± 7 | 207 ± 30 | 306 ± 31 | $2\times10^{-4}$ |
| | 1 cycle | $R_e(R_1Q_1)(R_2Q_2)Q_3$ | 37 ± 4 | 45 ± 6 | 82 ± 7 | $8\times10^{-5}$ |
| | 5 cycles | $R_e(R_1Q_1)(R_2Q_2)Q_3$ | 30 ± 2 | 50 ± 3 | 79 ± 4 | $7\times10^{-5}$ |
| | 10 cycles | $R_e(R_1Q_1)(R_2Q_2)Q_3$ | 28 ± 3 | 60 ± 4 | 88 ± 5 | $7\times10^{-5}$ |
| TREGDME_HCE | OCV | $R_e(R_1Q_1)(R_2Q_2)Q_3$ | 103 ± 10 | 42 ± 9 | 145 ± 13 | $1\times10^{-5}$ |
| | 1 cycle | $R_e(R_1Q_1)(R_2Q_2)Q_3$ | 42 ± 1 | 15 ± 2 | 56 ± 2 | $4\times10^{-5}$ |
| | 5 cycles | $R_e(R_1Q_1)(R_2Q_2)Q_3$ | 45 ± 1 | 25 ± 1 | 70 ± 1 | $3\times10^{-5}$ |
| | 10 cycles | $R_e(R_1Q_1)(R_2Q_2)Q_3$ | 47 ± 1 | 32 ± 1 | 79 ± 1 | $2\times10^{-5}$ |

The DEGDME_HCE and TREGDME_HCE are subsequently employed in a Li/LFP cell and galvanostatically cycled at the constant current rate of C/5 (1C = 170 mA g$^{-1}$) at the room temperature with the outcomes displayed in Figure 4. The voltage profiles reported in Fig. 4a (DEGDME_HCE) and Fig. 4c (TREGDME_HCE) reflect the response associated to the LiFePO$_4$ ⇌ Li + FePO$_4$ electrochemical process, centered at about 3.45 V as already described in CVs of Figure 3.[27,37] The two electrolytes show relatively limited polarization during the first cycles and capacity values approaching 160 mAh g$^{-1}$, that is, about 94 % of the theoretical value. Nonetheless, the coulombic efficiency interestingly approaches 100 % during both tests displayed by the cycling trends in Fig. 4b for DEGDME_HCE and in Fig. 4d for TREGDME_HCE, whilst a progressive increase of the polarization affects the cells after 10 cycles and leads to capacity decay upon 50 cycles, which is more remarkable for the former (Fig. 4a) compared to the latter electrolyte (Fig. 4c). Therefore, the cell using DEGDME_HCE (Fig. 4b) exhibits a capacity retention of 63 % during the 50 charge-discharge cycles taken into account, instead the one using TREGDME_HCE (Fig. 4d) holds 94 % of the initial capacity upon the same number of cycles. This behavior can be ascribed to the nature of the solid electrolyte interphase (SEI) layer formed at the electrodes surface upon cycling, which is influenced by the electrolyte composition, by the lithium salts content, and by the operating conditions.[38–40] In particular, the higher lithium salts concentration of TREGDME_HCE compared to DEGDME_HCE could play a crucial role in the formation of a more suitable SEI in this cell, as already suggested by its application in Li-$O_2$ cell studied elsewhere.[20] A further reason for the different Li/LFP cell performances between the two solutions may be the narrower electrochemical stability window of DEGDME_HCE (0 – 4.3 V) with respect to TREGDME_HCE (0 – 4.4 V).[20] However, the relevant increase of polarization observed for both DEGDME_HCE (Fig. 4a) and for TREGDME_HCE (Fig. 4c) after 50 cycles may actually indicate the need for further optimization of the SEI at the electrodes surface to allow a proper operation of the glyme-based electrolytes in a lithium-metal cell using insertion cathodes. Indeed, previous literature demonstrated that high concentrations of lithium salts in glyme-based electrolytes can lead to an uneven composition and low thickness of the SEI layer, possibly leading to a modest stability.[41] We have demonstrated in a previous paper that TREGDME dissolving lithium trifluoromethanesulfonate (LiCF$_3$SO$_3$) and LiNO$_3$ in conventional concentrations undergoes an electrochemical optimization process in Li/LiFePO$_4$ cell by adopting a reduction step at the first discharge occurring around 1.5 V, i.e., a voltage value far lower than the ones exploited in the galvanostatic measurements reported in Figure 4.[36] The above mentioned reduction deals with LiNO$_3$ and actually leads to the formation of stable interfaces at the electrodes surface with a remarkable improvement of the cell performance.[7,36]

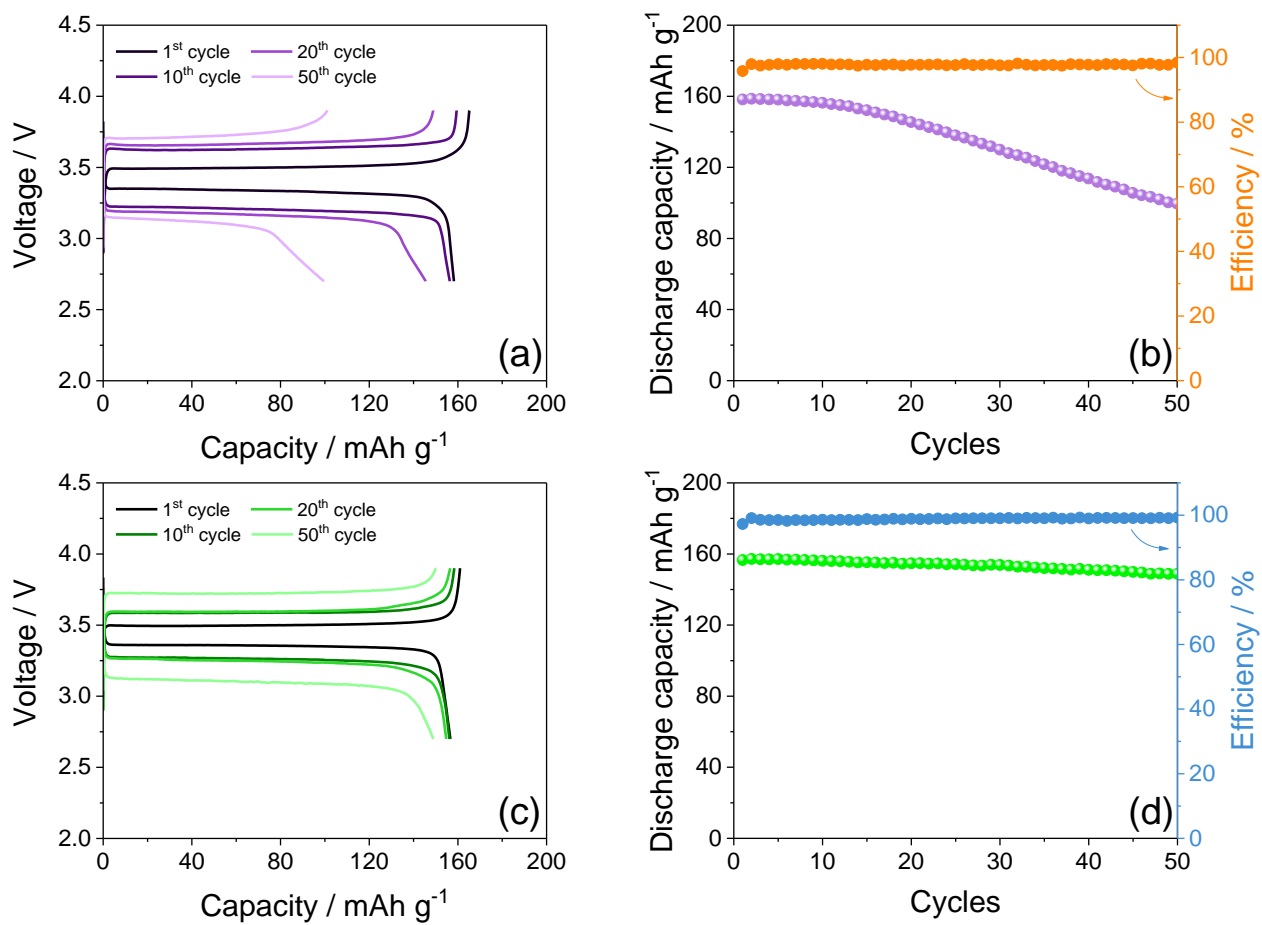

**Figure 4. (a,c)** Voltage profiles and **(b,d)** corresponding cycling trends with coulombic efficiency (right *y*-axis) related to Li/electrolyte/LFP cells employing either **(a,b)** DEGDME_HCE or **(c,d)** TREGDME_HCE galvanostatically cycled at the constant current rate of C/5 (1C = 170 mA g$^{-1}$) at room temperature (25 °C). Voltage range: 2.7 – 3.9 V.

Therefore, an additional galvanostatic test was performed using DEGDME_HCE and TREGDME_HCE in Li/LFP cells at the constant current rate of C/5 (1C = 170 mA g$^{-1}$) by using a voltage range between 1.2 and 3.9 V for the first cycle and between 2.7 and 3.9 V for the subsequent ones as reported in Figure 5. Insets in Fig. 5a (DEGDME_HCE) and Fig. 5c (TREGDME_HCE) show the voltage profile related to the first charge-discharge cycle, and reveal the evolution of a discharge plateau between 1.5 and 1.7 V due to the reduction of LiNO$_3$,[7,36] while the subsequent voltage profiles are reported in Fig. 5a and c, respectively. The cycling trends of the above Li/LFP cells evidence initial capacity values approaching 160 mAh g$^{-1}$ and coulombic efficiency around 100

% upon the first cycles for both DEGDME_HCE (Fig. 5b) and TREGDME_HCE (Fig. 5d). However, the cell employing DEGDME_HCE (Fig. 5b) still exhibits certain capacity decay, despite a retention increasing from 63 % to 71% compared to the analogue test performed without the additional reduction step (compare Fig. 4b and Fig. 5b) upon the 50 cycles taken into account. Furthermore, the voltage profiles of the cell using DEGDME_HCE (Fig. 5a) do not show the increase in cell polarization during cycling observed in the previous test (compare Fig. 4a and Fig. 5a), whilst a slope appears at the end of the charge and the discharge profiles after 20 cycles, and becomes more relevant after 50 cycles. The decrease of cell capacity and the appearance of the slope at the end of the (de-)insertion processes of LiFePO$_4$ may be associated with an excessive growth of the SEI layer at the electrodes during cycling using the DEGDME_HCE, and possibly with gradual changes in cathode crystallite size distribution and surface free energies of the lithiated and the de-lithiated phases of the olivine cathode which lead to a change of the biphasic potential.[11] Instead, the cell employing TREGDME_HCE shows a capacity retention increasing from 94 % of the previous test (Fig. 4d) up to 97 % (Fig. 5d), while the corresponding voltage profiles reveal only a slight slope after 50 cycles without any sign of polarization increase or deterioration (Fig. 5c). Therefore, we may suggest the additional reduction step at low voltage during the first cycle as an actual strategy to improve the performance of the lithium-metal cell using concentrated glyme-based electrolytes with LFP electrode, and particularly those having longer ether chain such as TREGDME.

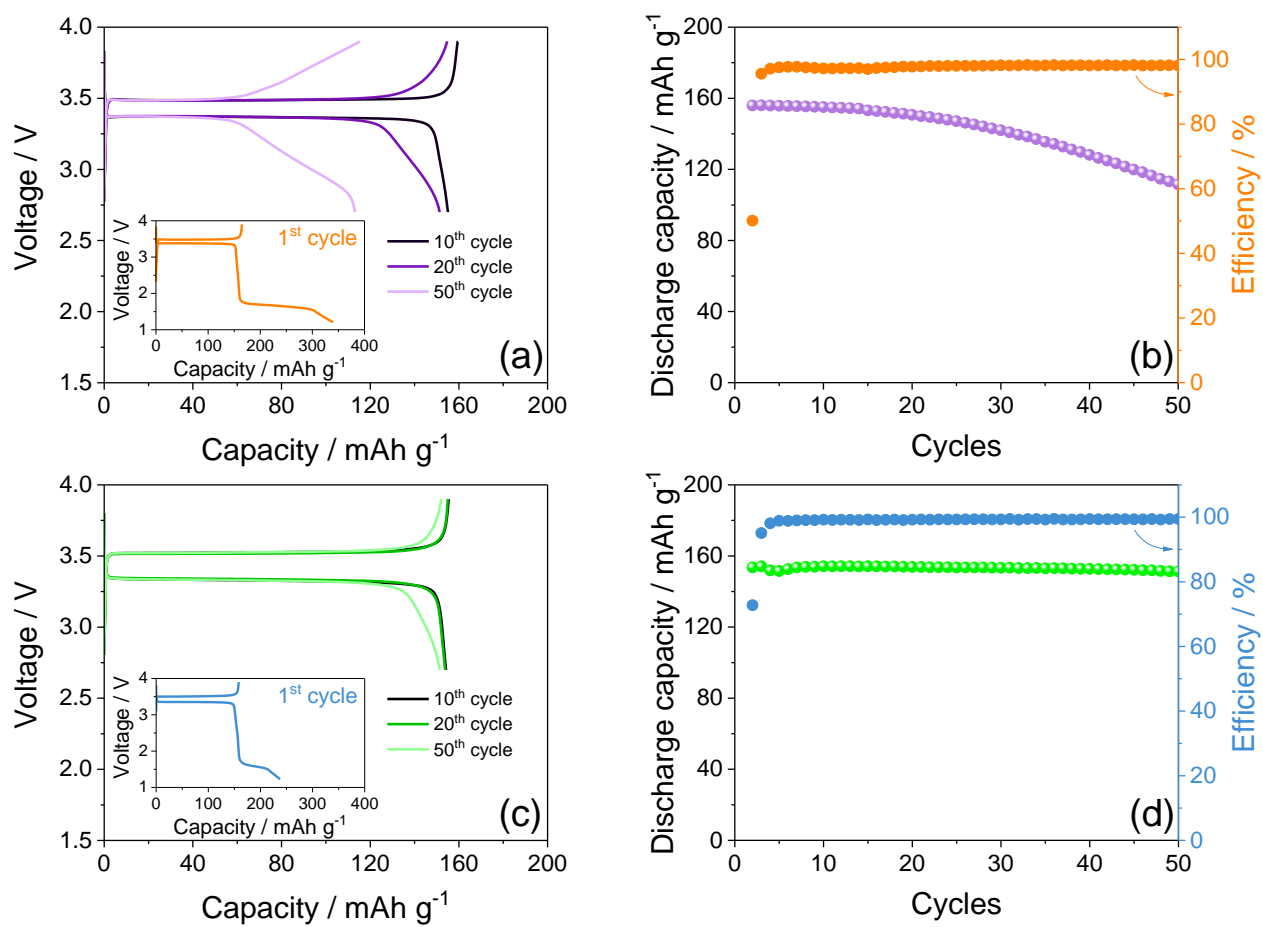

**Figure 5. (a,c)** Voltage profiles and **(b,d)** corresponding cycling trends with coulombic efficiency (right *y*-axis) related to Li/electrolyte/LFP cells employing either **(a,b)** DEGDME_HCE or **(c,d)** TREGDME_HCE galvanostatically cycled at room temperature (25 °C) at the constant current rate of C/5 (1C = 170 mA g$^{-1}$) by employing a voltage range between 1.2 and 3.9 V for the first cycle (inset in panels **(a)** and **(c)**) and between 2.7 and 3.9 V for the subsequent ones.

In order to extend the cycle life of the cells, additional galvanostatic cycling tests were performed on Li/S and Li/LFP cells by adopting the most suitable operative conditions according to the data reported in this work. Indeed, Figure 6 reports the cycling trends of Li/DEGDME_HCE/S:Sn 80:20 cell (Fig. 6a) operating at 35 °C and Li/TREGDME_HCE/LFP cell (Fig. 6b) working at room temperature (25 °C), the latter by exploiting the initial reduction step at low voltage (1.2 V). It is worth mentioning that the Li/S cell adopted an optimal electrolyte/sulfur ratio limited to 20 μl mgs$^{-1}$ in order to reduce the excess of electrolyte and, thus, to increase the practical energy density of the device.[30] As observed in Figure 6, both cells exhibit notable performances, long cycle life and coulombic efficiency around 100 % even by cycling at higher current rates, that is, at 1C for the Li/S cell (1C = 1675 mA $g_S^{-1}$) and C/3 for the Li/LFP one (1C = 170 mA $g^{-1}$). In particular, the Li/S cell delivers 140 cycles with an initial capacity upon activation of almost 750 mAh $g_S^{-1}$ retained at the 70 % at the end of the test (Fig. 6a), while the Li/LFP cell displays a capacity of 152 mAh $g^{-1}$ (89 % of the theoretical value) retained at the 85 % after 100 charge/discharge cycles (Fig. 6b). Despite the lower delivered capacity values with respect to the tests performed at C/5 (see Figs. 2 and 5), as expected by the employment of higher current rates, these tests further evidence that the optimal tuning of the working conditions can lead to the extension of the cycle life and to notable as well as steady delivered capacity values of Li/S and Li/LFP cells employing lowly flammable concentrated glyme-based electrolytes.

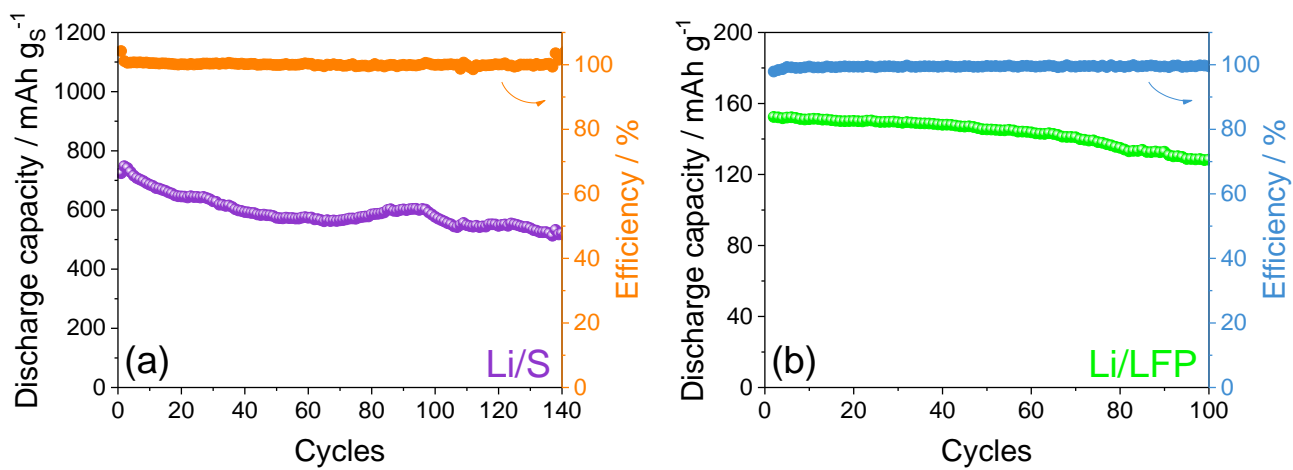

**Figure 6.** Cycling trends related to **(a)** Li/DEGDME_HCE/S:Sn 80:20 and **(b)** Li/TREGDME_HCE/LFP cells galvanostatically cycled at 1C (1C = 1675 mA $g_S^{-1}$) and C/3 (1C = 170 mA $g^{-1}$), respectively. Li/S cell was cycled at 35 °C by exploiting an electrolyte/sulfur ratio of 20 μl $mg_S^{-1}$ and a 1.6 – 2.8 V voltage range. Li/LFP cell was cycled at room temperature (25 °C) by employing a 1.2 – 3.9 V voltage range for the first cycle and voltage limits of 2.7 and 3.9 V for the following ones.

**Conclusions**

Glyme-based electrolytes with different chain length employing high lithium salts concentrations (indicated as DEGDME_HCE and TREGDME_HCE) are studied in either Li/S or Li/LFP cells to explore the applicability of this class of solutions in a safe and high-performances Li-metal battery. The CV tests performed on the Li/S cells revealed for both electrolytes a reversible electrochemical process centered at about 2.1 and 2.4 V, and an activation process leading to the decrease of the cell impedance from values of the order of 100/200 Ω to about 10 Ω upon the first CV cycle. Furthermore, galvanostatic measurements of the Li/S cells carried out using the constant current rate of C/5 at 25 and 35 °C indicated for DEGDME_HCE capacities of about 800 and 1300 mAh $g_S^{-1}$, respectively, while lower values of about 340 and 890 mAh $g_S^{-1}$ for TREGDME_HCE. The more relevant performances of the Li/S cells using DEGDME_HCE compared to TREGDME_HCE were attributed to a faster charge transfer kinetics in the former electrolyte compared to the latter. Meanwhile, the lithium cells employing the LFP electrode suggested the two solutions as possible electrolyte media for the reversible (de-)insertion process at about 3.45 V, however the tests indicated possible issues ascribed to the SEI layer formed at the electrodes surface, leading to polarization increase by cell cycling. These issues were relevantly mitigated, in particular using TREGDME_HCE, by adopting a first discharge of the cell extended down to 1.2 V in order to promote a further reduction of the $LiNO_3$ additive and the consolidation of a suitable SEI layer. Therefore, the above Li/LFP cells delivered at C/5 rate an initial capacity of about 160 mAh $g^{-1}$ (94 % of the theoretical value), an efficiency approaching 100%, and a capacity retention of 71% for DEGDME_HCE and 97% for TREGDME_HCE upon 50 charge/discharge cycles. Furthermore, Li/DEGDME_HCE/S and Li/TREGDME_HCE/LFP cells operating with the most adequate conditions have shown satisfactory performances at the high current rates of 1C and C/3, respectively. The former cell delivered 750 mAh $g_S^{-1}$ with capacity retention of 70 % over 140 cycles at 35 °C, while the latter exhibited about 150 mAh $g^{-1}$ with a retention of 85 % after 100 cycles at 25 °C by exploiting the initial reduction step at 1.2 V mentioned above.

In summary, the findings of this work suggested the possible use of concentrated solutions based on end-capped glymes in efficient lithium-metal cells by careful tuning of *i)* the ether chain length, *ii)* the salt nature and concentration, *iii)* the chemistry of the cathode material, and *iv)* the operative conditions, including temperature and voltage limits. In addition, the intrinsically lower flammability of the concentrated glymes reported herein compared to the common electrolytes used in battery is expected to allow the development of a lithium-metal battery with high energy and acceptable safety content.

**Acknowledgements**

This project/work has received funding from the European Union's Horizon 2020 research and innovation programme Graphene Flagship under grant agreement No 881603. The authors also thank grant "Fondo di Ateneo per la Ricerca Locale (FAR) 2019", University of Ferrara, and the collaboration project "Accordo di Collaborazione Quadro 2015" between University of Ferrara (Department of Chemical and Pharmaceutical Sciences) and Sapienza University of Rome (Department of Chemistry).

**Table of content image**

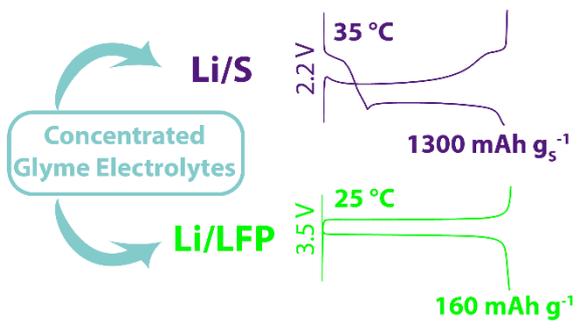